\documentclass[aps,prc,showpacs,amsmath,amssymb]{revtex4}
\usepackage{epsfig}

\begin{document}

\title{Eta electroproduction on nuclei in the nucleon resonance region}

\author{J. Lehr and U. Mosel}

\affiliation{Institut f\"ur Theoretische Physik, Universit\"at Giessen\\ 
            D-35392 Giessen, Germany}

\date{\today}

\begin{abstract}
We investigate eta electroproduction on nuclei for $Q^2=2.4$ and 3.6 GeV$^2$
in the framework of a coupled-channel BUU transport model. We analyze the 
importance of final 
state interactions and side feeding and compare with findings drawn from 
eta photoproduction. It is shown that in contrast to photoproduction the 
influence of etas stemming from secondary processes becomes important at 
high $Q^2$.
\end{abstract}
\maketitle

\section{Introduction}

Photon and electron induced reactions provide a useful tool to probe 
particle properties in the nuclear medium. In the nucleon resonance 
region, meson production can be used to learn about the in-medium properties 
of nucleon resonances. Among other processes, measurements of eta 
photoproduction were performed \cite{krusche_eta,yorita_eta} in order to 
obtain information
about the $S_{11}(1535)$. This channel is particularly interesting because
of the strong coupling between the eta and the $S_{11}(1535)$ in the energy 
region from the $\eta N$ threshold up to invariant masses of about 1.6 GeV.
In photoproduction the momentum transfer and the mass of the created 
resonance are directly related to each other: In the vacuum, a $S_{11}$ with
pole-mass of $\mu=1.535$ GeV can only be excited by a real photon with
energy $E_\gamma\sim 0.787$ GeV; inside the nucleus this condition is 
somewhat softened due to Fermi motion. 

Only recently, we have calculated $\eta$ photoproduction on different
nuclei in the region of the $S_{11}(1535)$ \cite{photo_eta}. The agreement
with the available data is good when a momentum dependent resonance potential
is applied. We have found collisional broadening for the $S_{11}$ of only 
about 30 MeV that had almost no visible effect on the cross section. It is 
therefore interesting to see whether this holds over a larger range of 
resonance momenta. For a review of photon induced processes within our model 
we refer to \cite{effe_pion,lehr_electro,photo_eta}.

In electroproduction experiments, the electron interacts with the nucleus by 
the exchange of a single virtual photon (one-photon exchange approximation). 
This yields an
additional degree of freedom, because photon energy and momentum can now be
chosen independently from each other. Therefore, the resonance self energy
can be probed over a full range of mass and momentum.
Electroproduction of mesons in the resonance region at low 
$Q^2\le 0.8$ GeV$^2$ was already addressed in the framework of the 
coupled-channel BUU model in \cite{lehr_electro}.

A disadvantage of electron induced processes is the decrease of the 
cross sections with increasing $Q^2$. Especially the resonant structure
becomes less pronounced due to the $Q^2$ dependence of the photocoupling 
helicity amplitudes (see e.g. \cite{stoler}). In addition, Fermi motion 
becomes the more effective the larger $Q^2$ is \cite{lehr_electro}, resulting
in a smearing of the resonant structure over a wide energy range and an
even stronger decrease of the cross section maxima.

In this work we calculate eta electroproduction on nuclei within a
semi-classical coupled-channel BUU transport model; the same model was used 
in \cite{photo_eta}
to calculate photon induced eta production. In this work, we investigate eta
production mechanisms in electroproduction in the kinematical regime
accessible at JLab and compare with findings from eta photoproduction.
This makes it possible to study the transition from low resonance momenta
at $Q^2=0$ to larger momenta at $Q^2=3.6$ GeV$^2$, i.e. probing the momentum 
regime of a pole-mass resonance from 0.8 GeV up to more than 3 GeV.

In Sec. \ref{sec:model} we briefly discuss the BUU model, our 
approach to the $\gamma^* N\to \eta N$ reaction and the treatment of
the final state interactions (FSI). In Sec. \ref{sec:results} we 
qualitatively compare differences between eta photo- and electroproduction.

\section{The Model}
 \label{sec:model}

We treat the electron-nucleus reaction in the one-photon exchange
approximation. Therefore, the process can be described in the same way as in
photoproduction. The model then contains two aspects: the reaction
of the (virtual) photon with the nucleus and the treatment of the final state
interactions (FSI).

\subsection{The photon-nucleon reaction}

For the elementary reaction, we assume that the photon is absorbed by a 
single nucleon. In the nucleon resonance region the relevant processes
are 
\begin{equation}
  \label{eq:elem_reactions}
  \gamma^* N\to P_{33}(1232),\quad \gamma^* N\to S_{11}(1535),\quad
  \gamma^* N\to F_{15}(1680),\quad \gamma^* N\to N\pi,\quad 
  \gamma^* N\to N\pi\pi.
\end{equation}
For photoproduction, also the $D_{13}(1520)$ has to be taken into account 
\cite{effe_pion}, but the helicity amplitude of this resonance decreases 
rather quickly with increasing $Q^2$ \cite{stoler}, especially compared to 
the $S_{11}$, so that its contribution to the second 
resonance region becomes small for $Q^2\sim 3$ GeV$^2$. Moreover, the
$D_{13}$ couples only very weakly, if at all, to the eta 
\cite{pdg,penner}.
As an input for our model, we need cross sections for the different
channels displayed in (\ref{eq:elem_reactions}). Unlike the case of
photoproduction, where a large body of data can be exploited \cite{effe_pion},
at finite $Q^2$ only few processes have been measured. We present
calculations for $Q^2=2.4$ and 3.6 GeV$^2$, because we then can use the
recently measured cross sections for the primary reaction
$\gamma^* p\to\eta p$ by Armstrong
\emph{et al.} \cite{armstrong_eta}. We assume that in the considered
energy regime the reaction proceeds by the excitation and subsequent decay
of a $S_{11}(1535)$ resonance; background contributions are negligible
\cite{armstrong_eta}. Therefore, we use a Breit-Wigner parametrization 
similar to that applied in eta photoproduction \cite{photo_eta}:
\begin{equation}
  \label{eq:gamma_proton_eta}
  \sigma_{\gamma p\to \eta p}=\left({k_0\over k}\right)^2
   {s\Gamma_\gamma(\sqrt s)\Gamma_{S_{11}\to \eta p}(\sqrt s)\over
   (s-M_{S_{11}}^2)^2+s\Gamma_{S_{11}\to X}^2(\sqrt s)} 
   {2m_N\over M_{S_{11}}}\vert A_T\vert^2,
\end{equation}
with $\Gamma_\gamma=k/k_0$ and the total and partial resonance widths 
$\Gamma_{S_{11}\to X}(\sqrt s)$ and $\Gamma_{S_{11}\to\eta N}(\sqrt s)$. 
For the latter two we use the parametrizations from \cite{manley,effe_dilep}.
$k=k(\sqrt s)$ denotes the photon center of mass (cm) momentum 
and $k_0=k(M_{S_{11}})$, where $M_{S_{11}}$ is the resonance pole mass.
The transversal photocoupling helicity amplitude $A_T$ is determined from the 
fit to the exclusive data in \cite{armstrong_eta}; it thus contains the
form factor of the resonance. The longitudinal
contribution to the resonant cross section is neglected, which is a
reasonable assumption \cite{stoler}.
The obtained values are shown in Tab. \ref{tab:helicity-ampl} and are in
line with the findings in \cite{armstrong_eta}. 

\begin{figure}
\begin{center}
\includegraphics[width=11cm]{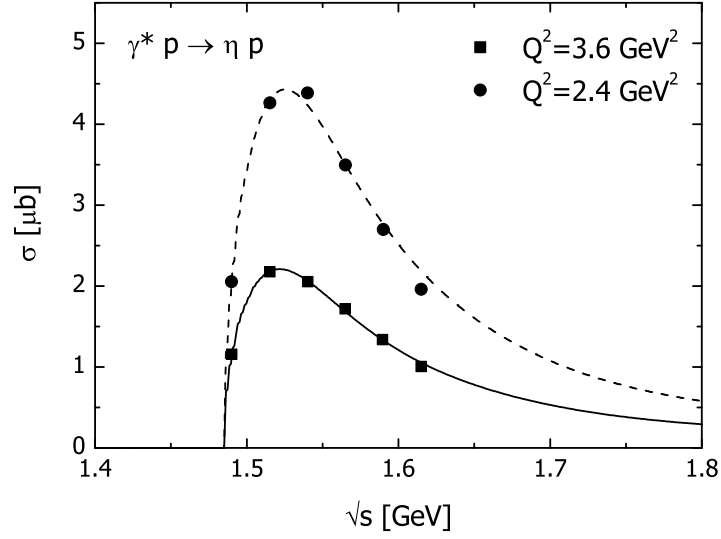}
\end{center}
\vspace{-0.5cm}
\caption{Parametrization of the elementary process $\gamma^* p\to\eta p$ 
for $Q^2=2.4$ and 3.6 GeV$^2$ according to Eq. 
(\protect\ref{eq:gamma_proton_eta}).
The data are taken from \protect\cite{armstrong_eta}.}
 \label{fig:gamma_proton}
\end{figure}

In Fig. 
\ref{fig:gamma_proton} we show the parametrization in comparison with the 
data. The agreement is very good. Note that the energy dependence of the 
curves is strongly influenced by the energy parametrization of the resonance 
widths. 

It is not sufficient to account only for this direct process, because via
the FSI all other elementary channels may contribute to eta production.
Due to the absence of data for the other exclusive channels in 
(\ref{eq:elem_reactions}) it is difficult to determine the 
respective contributions unambiguously. For the total cross section 
$\gamma^* p\to X$
we make use of the data compilation by Brasse \emph{et al.} \cite{brasse76}. 
The cross section $\gamma^* p\to S_{11}\to X$ can be obtained 
from (\ref{eq:gamma_proton_eta}) by substituting the partial decay width in the
numerator with the total width: 
\begin{equation}
  \label{eq:gamma_proton_tot}
  \sigma_{\gamma^* p\to S_{11}\to X}=\sigma_{\gamma^* p\to S_{11}\to\eta p}
  \cdot{\Gamma_{S_{11}\to X}\over \Gamma_{S_{11}\to\eta p}}.
\end{equation}
The remaining resonance contributions of the $P_{33}(1232)$ and the  
$F_{15}(1680)$ are determined by making Breit-Wigner parametrizations 
according to Eq. (\ref{eq:gamma_proton_eta}). The used transversal helicity
amplitudes are also displayed in Tab. \ref{tab:helicity-ampl}.
The $P_{33}$ amplitudes are close to those cited in \cite{stoler}. For the
$F_{15}$ much larger values (factor of $\sim 2$) were necessary in order to 
saturate the whole
resonant structure in the third resonance region suggested by the inclusive
data in \cite{brasse76}. The background contributions $\gamma^* p\to N\pi$
and $N\pi\pi$ were fixed by absorbing the remaining strength after summing 
all contributions incoherently to the total cross section.

In the neutron sector, there is no experimental
information at all. Therefore, we use the same cross sections as for the
protons, except for the channel $\gamma^* n\to S_{11}$. Here we use, as in
photoproduction, the relation $\sigma_n=2/3\sigma_p$ 
\cite{photo_eta,krusche_deut}.
 
We want to stress that the data situation does not allow a more refined 
separation 
of the total cross section into the different channels. On the other hand, the
very important primary eta source $\gamma^* p\to S_{11}\to\eta p$ is well 
under control and the total cross section also is described by our 
parametrization. Moreover, the details of the separation of strength into the 
different pionic channels do not influence the more qualitative issues we 
will discuss in Sect. \ref{sec:results}.
\bigskip

The vacuum cross sections $\sigma_{\gamma^* N}$ discussed so far are
connected to the experimentally measured cross sections for the 
electron-nucleon reaction. Using the Hand convention \cite{hand}, we have
\begin{equation}
  \label{eq:electro-xsec}
  {d\sigma\over d\Omega dE^\prime}=\Gamma\cdot\sigma_{\gamma^* N}
  ={\alpha\over 2\pi^2}{E^\prime\over E}{k_\gamma\over Q^2}
   {1\over 1-\varepsilon}\cdot\sigma_{\gamma^* N}.
\end{equation}
Here $E$, $E^\prime$ and $\varepsilon$ are the energies of the incoming
and outgoing electrons and the degree of longitudinal polarization of the
virtual photons, respectively. $\Omega$ refers to the scattering angle 
of the electrons in the lab frame and $k_\gamma=(s-m_N^2)/(2m_N)$ denotes the
equivalent photon momentum, where $\sqrt s$ is the cm energy of the 
$\gamma^* N$ pair. 

In order to measure the excitation function of the $S_{11}(1535)$ as a function
of the photon energy at fixed 
$E$ and $Q^2$, one has to vary the electron scattering angle and the energy 
of identified outgoing electrons.
The eta data on the proton \cite{armstrong_eta} were obtained at fixed electron
scattering angle, so that actually a certain finite $Q^2$ range was covered. 
This was accounted for by correcting the data accordingly. 
We use electron energies $E=3.2$ and 4 GeV for $Q^2=2.4$
and 3.6 GeV$^2$, respectively, as in \cite{armstrong_eta}.
\bigskip

The nucleons are distributed in the nucleus according to a Woods-Saxon
distribution. We apply the local density approximation with local Fermi 
momenta 
to generate the momentum distribution if the nucleons. For the binding of the 
nucleons we use a density and momentum dependent potential described in 
\cite{photo_eta}.
As a result, the kinematics of the nucleons and hence the cm energy of the
initial photon-nucleon pairs differ from the vacuum case.

The kinematical situation in the considered processes is such that 
shadowing effects in the photon-nucleus reaction are negligible.

\subsection{The BUU model and final state interactions}
 \label{sec:fsi}

The aspect of the FSI is treated within a coupled-channel BUU transport model,
which is based upon the BUU equation. This equation describes the evolution 
of the phase space density $F_i$ of a certain particle type $i$:
\begin{equation}
  \label{eq:buu-eq}
  \left({\partial\over\partial t}+\vec\nabla_p H\cdot\vec\nabla_r
  -\vec\nabla_r H\cdot\vec\nabla_p\right)F_i(\vec r,\vec p,\mu;t)
= I_{\textrm{coll}}[F_N,F_\pi,F_{P_{33}(1232)},F_\eta,...]
\end{equation}
The left-hand side describes the particle propagation under the influence
of a Hamilton function $H=\sqrt{(\mu_i+S_i)^2+p_i^2}$, which in the case
of baryons includes an effective scalar potential $S_i$. Besides the nucleon, 
our model contains nucleon resonances and the relevant mesonic degrees of 
freedom $\pi$, $\eta$, $\rho$, ... We use the set of 29 nucleon resonances and
parameters from Manley and Saleski \cite{manley}. 
The collisional integral on the right-hand side accounts for the coupling to 
the phase space densities of other particles due to reactions such as 
collisions, decay and resonance formation. The collision reactions are usually
of binary type. 
The eta meson couples to the three resonances $S_{11}(1535)$, $S_{11}(1650)$
and $F_{17}(1990)$. As already described above, we only account for the first 
of these resonances in the $\gamma^* N$ reaction. Relevant FSI for eta 
production may be
$NR\leftrightarrow NN$, $NR\leftrightarrow NR^\prime$, $m N\leftrightarrow R$.

The reactions at large $Q^2$ with large momentum transfer in the elementary 
reaction involve collisions in the FSI partially with invariant masses
above 2 GeV. 
For such reactions we use the string model Fritiof \cite{fritiof} to determine
the particle content of the final states and the kinematics.
Relevant final states in the considered energy range contain mainly the
$\eta$ and $\eta^\prime$ along with nucleons, $\Delta$ resonances and pions.
For more model details we refer to \cite{photo_eta,effe_dilep,effe_cebaf}.

\section{Results}
 \label{sec:results}

We now present our results for the reaction $e\,^{40}\textrm{Ca}\to e^\prime
\eta X$ for $Q^2=2.4$ and 3.6 GeV$^2$ and compare with eta photoproduction
$\gamma\, ^{40}\textrm{Ca}\to\eta X$. We do not take into account any medium 
modifications for resonances.

\begin{figure}
\begin{center}
\includegraphics[width=13cm]{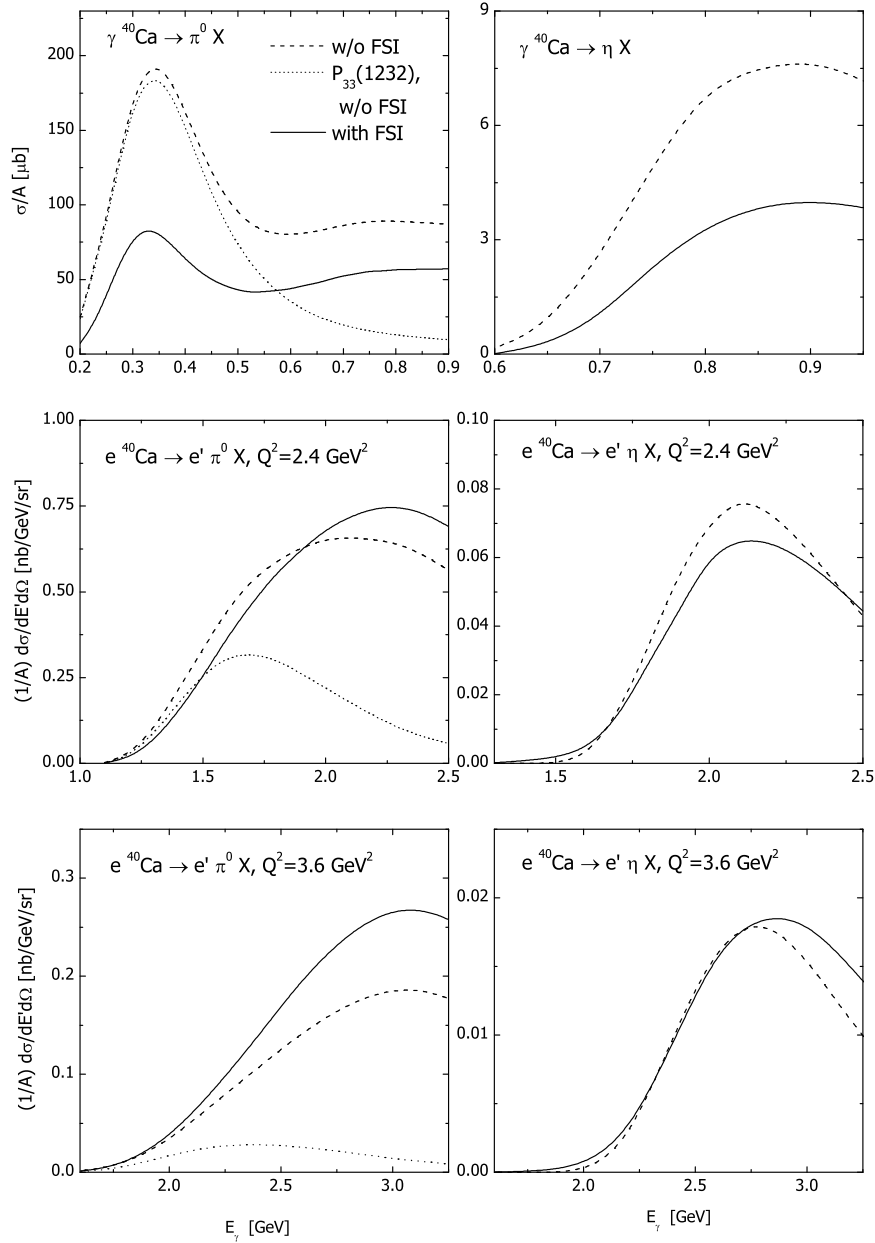}
\end{center}
\vspace{-0.5cm}
\caption{Pion and eta production on Calcium. The top panels show 
photoproduction results, the lower ones results for electroproduction 
at the given $Q^2$ values. The solid and dashed lines show calculations with
and without FSI, respectively. The dotted lines visualize the contribution of 
the $P_{33}(1232)$ to the cross sections without FSI.}
 \label{fig:fsi_nofsi}
\end{figure}

\begin{figure}
\begin{center}
\includegraphics[width=11cm]{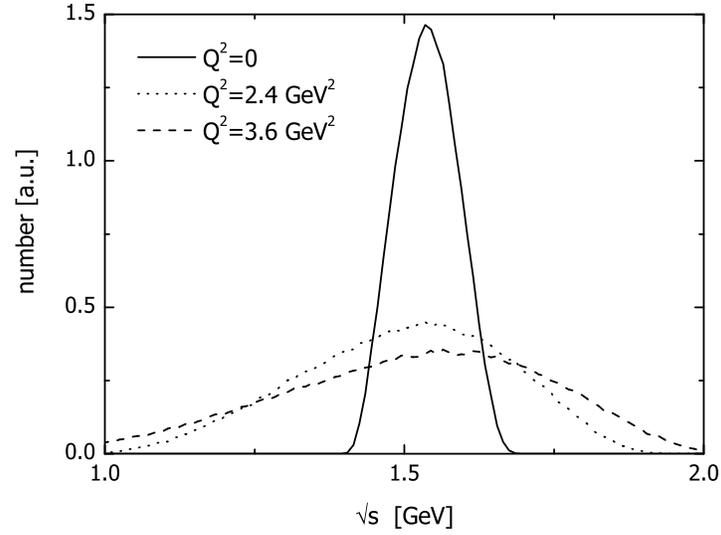}
\end{center}
\vspace{-0.5cm}
\caption{$\sqrt s$ spectra of the elementary $\gamma N$ pairs for 
different $Q^2$. The respective photon energies were chosen such that a
resonance with mass $\mu=\sqrt s=1.535$ GeV is excited on a nucleon at rest.}
 \label{fig:srts_spec}
\end{figure}

In Fig. \ref{fig:fsi_nofsi} we discuss the influence of the FSI on pion and
eta photo- and electroproduction. 
First, we focus on the calculations obtained without FSI (dashed curves). 
That is, only the effects of Fermi motion, Pauli blocking and nuclear binding 
on the elementary reaction are studied. In the case of pions (left panels) the
whole resonance region is shown. It is seen that the resonant structure, 
clearly visible in photoproduction, is totally washed out at larger values of 
$Q^2$. In particular, the peak of the first resonance region has  
disappeared. The remaining, flat contribution of the $P_{33}(1232)$ is shown 
as dotted curves.
The behavior is caused by the strong decrease of the resonance helicity
amplitudes compared to the background processes and the more effective
Fermi smearing at large $Q^2$. This is demonstrated in Fig. 
\ref{fig:srts_spec},
where the the cm energy of the primary photon-nucleon pair is shown for
different values of $Q^2$. Here the photon energy in each case is chosen such 
that a cm energy of $M_{S_{11}}=1.535$ GeV is obtained for a nucleon at rest.
It is seen that the spectrum becomes very broad for $Q^2=3.6$ GeV$^2$,
covering actually the whole resonance region, wheras the spectrum at $Q^2=0$
is confined to the second resonance region. Therefore, at fixed photon energy 
a much wider cm energy range is probed than in photoproduction.

In eta production, the resonance peak structure of the $S_{11}(1535)$ is
visible in Fig. \ref{fig:fsi_nofsi} for each $Q^2$ value, because no other 
processes contribute to this channel in the absence of FSI. 
\bigskip

When the FSI are turned on (solid curves), we observe at $Q^2=0$, that the
curves drop for both pions and etas: The particles that stem from the 
elementary reaction can be absorbed. In the pionic channel this happens
by collisional reactions such as $NR\to NN$ (see 
\cite{effe_pion,lehr_electro}), whereas in the eta channel this is 
only of small importance \cite{photo_eta}. Here the main absorption 
mechanism is $\eta N\to R\to \pi N$.

With increasing $Q^2$, the picture changes drastically: The curves for the 
pions are larger than the results without FSI for photon energies 
$\gtrsim 2$ GeV. For $Q^2=3.6$ GeV$^2$ this happens already at the beginning
of the first resonance region. The reason is that the particles stemming from
the elementary reaction (\ref{eq:elem_reactions}) are produced at larger
momentum transfer ($q_\gamma\sim 2.6$ GeV for $Q^2=2.4$ GeV$^2$ and
$q_\gamma\sim 3.3$ GeV for $Q^2=3.6$ GeV$^2$ on top of the $S_{11}$
resonance peak). Therefore the background pions and decay products of the 
resonances are capable of producing even more particles in the FSI. This 
effect is missing in photoproduction, where the pion kinetic energies are not 
large enough at lower photon energies. The result is that the particle loss
by absorption, which is of course also present at finite $Q^2$, becomes
dominant.

In eta production we observe the same effect. For the higher $Q^2$ values
the curves with and without FSI lie essentially on top of each other
in the threshold region. For 
$Q^2=3.6$ GeV$^2$ the effect is even more pronounced.
Note also that at finite $Q^2$ the cross section close to the threshold 
is larger than in the calculations without FSI due to secondary
processes. Such contributions are also absent in photoproduction.
\bigskip

\begin{figure}
\begin{center}
\includegraphics[width=15cm]{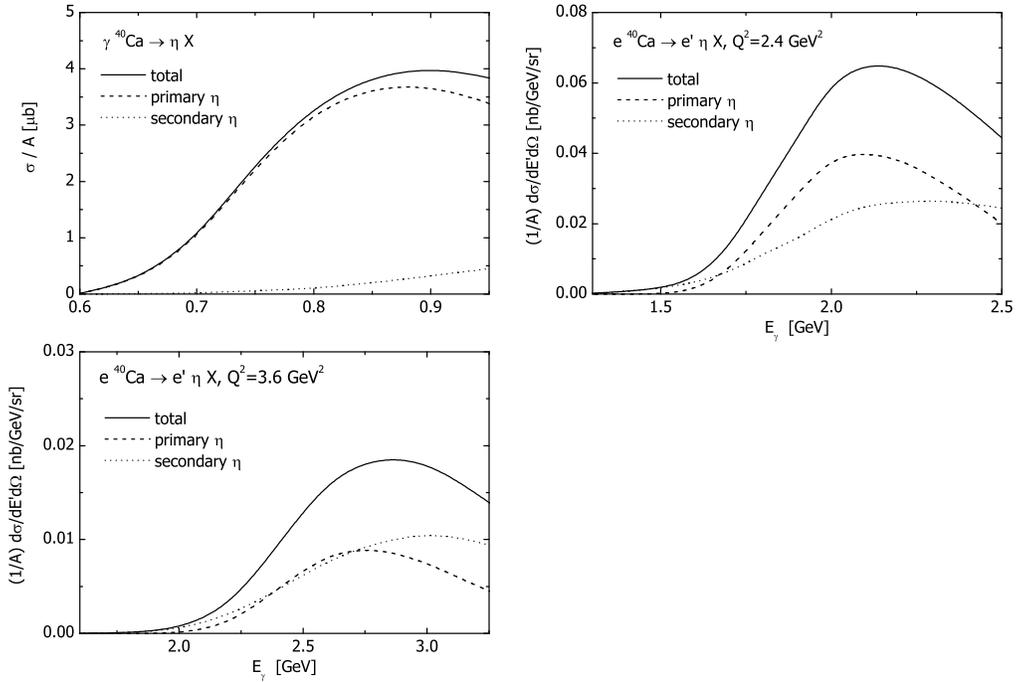}
\end{center}
\vspace{-0.5cm}
\caption{Contributions from primary and secondary etas to the cross section
on Calcium for different values of $Q^2$.} 
 \label{fig:prim_sek}
\end{figure}

In Fig. \ref{fig:prim_sek} we analyze the origin of the observed etas again 
for photo- and electroproduction. It is seen that with increasing $Q^2$ the 
relative importance of
the primarily produced etas (from the reaction $\gamma N\to S_{11}\to \eta N$)
decreases. This is partially caused by the decrease of the helicity
amplitudes and the Fermi smearing already mentioned. On the other
hand, the contribution of secondary etas produced in the FSI
increase and become more important for $Q^2=3.6$ GeV$^2$.
It is
remarkable that the production mechanism changes to such a degree. This also
influences the information content of eta production with respect to the 
$S_{11}(1535)$ properties in medium, because one is less sensitive to the 
resonances produced in the first reaction.
\bigskip

\begin{figure}
\begin{center}
\includegraphics[width=15cm]{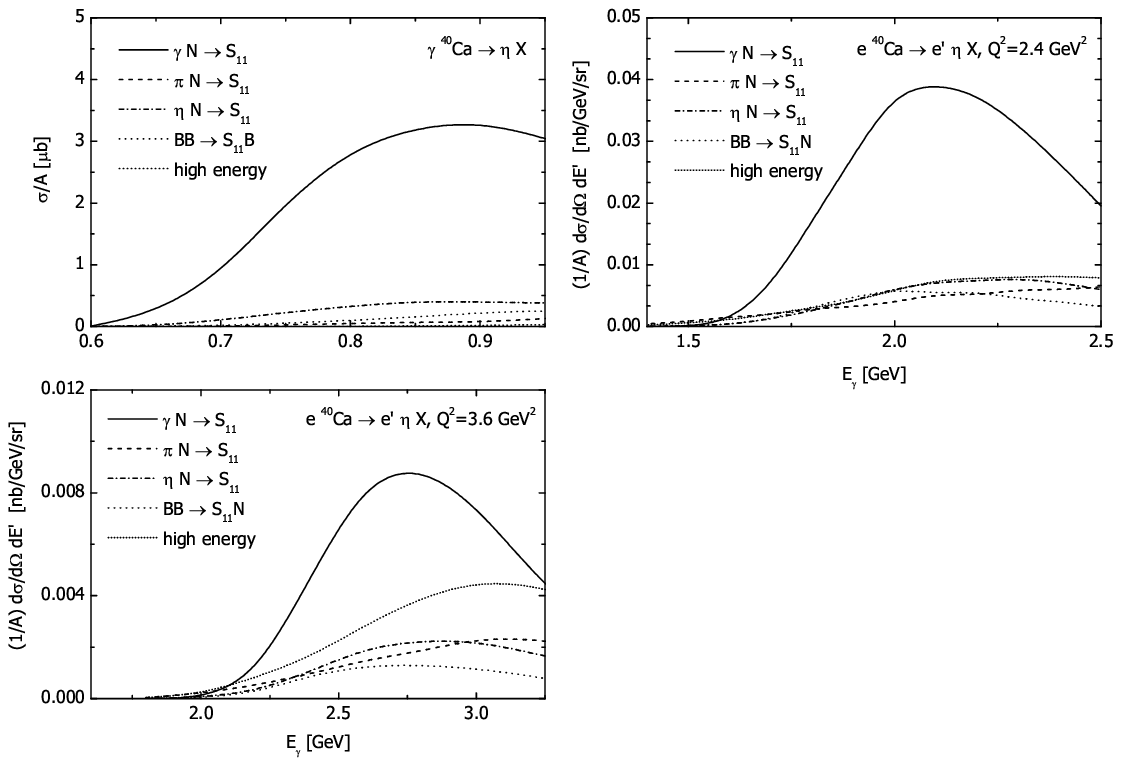}
\end{center}
\vspace{-0.5cm}
\caption{Contributing channels for eta photo- and electroproduction on
Calcium. The detected etas stem either from the decay of $S_{11}$ resonances
produced in the reactions given in the legend or from high energy reactions.
The solid line in each case is the same as the dashed curve in Fig.
\ref{fig:prim_sek}, whereas the other curves depict the secondary production
channels.} 
 \label{fig:contributions}
\end{figure}

In Fig. \ref{fig:contributions} we discuss the different sources the detected 
eta actually stem from. The processes $\gamma N$, $\pi N$,
$\eta N\to S_{11}$ and $BB\to S_{11} N$ denote etas resulting from the decay 
of $S_{11}$ resonances that were produced in such reactions. The 'high energy'
contributions denote etas resulting from baryon-baryon or baryon-meson 
collisions at larger energies determined by the Fritiof model either directly 
or by the decay of $S_{11}$ and $\eta^\prime$ resonances produced in such 
events or elastic scattering at higher energies. We see that the main 
contribution still consists of etas that did not get absorbed and re-emitted 
after their first production via the $S_{11}$ decay. However, as already 
described above, at higher $Q^2$ secondary processes become more
important, especially the high energy contributions and reactions from
$\pi N,\eta N\to S_{11}$. The latter are also responsible for the subthreshold 
contributions for energies below the threshold of the $\gamma N\to S_{11}$ 
channel.

\section{Summary}

We have calculated eta production in photon and electron induced
reactions on nuclei for several values of $Q^2$. This is of particular 
interest, because at large $Q^2$ the physics of the resonance region (i.e. 
reactions with invariant masses below 2 GeV) is mixed with FSI at larger 
energies than encountered in photoproduction. 

We have found that the FSI mechanisms change with increasing $Q^2$ from an 
essentially absorptive character at $Q^2=0$ to a source of relevant 
contributions. This is due to the increasing momentum transfer in the 
elementary reaction and an opening of secondary eta production channels in 
the FSI. 
It is clear that the inclusion of such side-feeding effects is important for 
the interpretation of eta electroproduction in terms of possible in-medium 
modifications of the $S_{11}(1535)$ resonance. 
Furthermore, we have found that the FSI also lead to an increase of the cross
sections for pion and eta production.
A similar effect has been seen in high-energy photo production of $K^+$ 
\cite{effe_cebaf} and of $\rho$ mesons \cite{falter_rho} on nuclei.

\section*{ACKNOWLEDGMENTS}

This work was supported by DFG.


\newpage

\begin{table}
\caption{\label{tab:helicity-ampl} Transversal helicity amplitudes $A_T$ for 
the resonances.}
\begin{ruledtabular}
\begin{tabular}{lccc}
 $Q^2$ [GeV$^2$] & $A_T(P_{33}(1232))$ [GeV$^{-1/2}$] 
                 & $A_T(S_{11}(1535))$ [GeV$^{-1/2}$] 
                 & $A_T(F_{15}(1680))$ [GeV$^{-1/2}$] \\
 \hline
 2.4 & 0.0688 & 0.0569 & 0.0674   \\
 3.6 & 0.0316 & 0.0401 & 0.0512   \\
\end{tabular}
\end{ruledtabular}
\end{table}

\begin{thebibliography}{99}

\bibitem{krusche_eta} B. Krusche \emph{et al.}, Phys. Rev. Lett. {\bf 74}, 
    3736 (1995).
\bibitem{yorita_eta} T. Yorita \emph{et al.}, Phys. Lett. {\bf B476}, 226 
   (2000).
\bibitem{photo_eta} J. Lehr, M. Post, U. Mosel, nucl-th/0306024,
  submitted to Phys. Rev. C
\bibitem{effe_pion} M. Effenberger, A. Hombach, S. Teis, U. Mosel,
  Nucl. Phys. {\bf A614}, 501 (1997).
\bibitem{lehr_electro} J. Lehr, M. Effenberger, U. Mosel, Nucl. Phys.
{\bf A671}, 503 (2000).
\bibitem{stoler} P. Stoler, Phys. Rep. {\bf 226}, 103 (1993).
\bibitem{pdg} Particle Data Group, Review of Particle Physics,
   K. Hagiwara \textit{et al.}, Phys. Rev. \textbf{D66}, 010001 (2002).
\bibitem{penner} G. Penner, U. Mosel, Phys. Rev {\bf C66}, 055212 (2002).
\bibitem{armstrong_eta} C.S. Armstrong \emph{et al.}, Phys. Rev. {\bf D60},
    052004 (1999).
\bibitem{manley} D.M. Manley, E.M. Saleski, Phys. Rev. {\bf D45}, 4002  
  (1992).
\bibitem{effe_dilep} M. Effenberger, E.L. Bratkovskaya, U. Mosel, Phys. Rev.
{\bf C60}, 044614 (1999).
\bibitem{brasse76} F.W. Brasse, W. Flauger, J. Gayler, S.P. Goel, R. Haidan,
 M. Merkwitz, H. Wriedt, Nucl. Phys. {\bf B110}, 413 (1976).
\bibitem{krusche_deut} B. Krusche \emph{et al.}, Phys. Lett. {\bf B358}, 
   40 (1995).
\bibitem{hand} L.N. Hand, Phys. Rev {\bf 129}, 1834 (1963).
\bibitem{fritiof} B. Andersen, G. Gustafson, Hong Pi,
    Z. Phys. \textbf{C57}, 485 (1993).
\bibitem{effe_cebaf} M. Effenberger, U. Mosel,
Phys. Rev. {\bf C62}, 014605 (2000).
\bibitem{falter_rho} T. Falter, K. Gallmeister, U. Mosel,
Phys. Rev. {\bf C67}, 054606 (2003).
\end{thebibliography}
\end{document}